\documentclass[journal=jpclcd,manuscript=article]{achemso}

\usepackage[version=3]{mhchem} 
\usepackage{color}
\usepackage{xspace}
\usepackage{geometry}
\usepackage{graphicx}
\usepackage{float}
\usepackage{chemfig}
\usepackage{subcaption}
\usepackage{adjustbox}
\usepackage{amsfonts}
\usepackage{etoolbox}
\usepackage{hyperref}
\usepackage{pgf}

\usepackage[authormarkuptext=name,authormarkup=none]{changes}
\definecolor{blue}{RGB}{0, 0, 255}
\definechangesauthor[name={Yorick}, color={blue}]{YS}
\definechangesauthor[name={Ivan}, color={violet}]{IT}
\definechangesauthor[name={Tetiana}, color={magenta}]{TO}

\usepackage[normalem]{ulem} 

\pgfkeys{
  /molecules/0/.initial=\ce{MPF}\xspace,
  /molecules/3/.initial=\ce{tBu{-}MPF}\xspace,
  /molecules/10/.initial=\ce{FtBu{-}MPF}\xspace,
  /molecules/7/.initial=\ce{TIF}\xspace,
  /molecules/9/.initial=\ce{tBu{-}TIF}\xspace,
  /molecules/11/.initial=\ce{FtBu{-}TIF}\xspace,
  /molecules/49/.initial=\ce{FYT}\xspace,
  /molecules/94/.initial=\ce{tBu{-}FYT}\xspace,
  /molecules/99/.initial=\ce{FtBu{-}FYT}\xspace,
  /molecules/50/.initial=\ce{MMPIF}\xspace,
  /molecules/95/.initial=\ce{tBu{-}MMPIF}\xspace,
  /molecules/100/.initial=\ce{FtBu{-}MMPIF}\xspace,
  /molecules/60/.initial=\ce{MXT}\xspace,
  /molecules/96/.initial=\ce{tBu{-}MXT}\xspace,
  /molecules/101/.initial=\ce{FtBu{-}MXT}\xspace,
  /molecules/68/.initial=\ce{FMT}\xspace,
  /molecules/97/.initial=\ce{tBu{-}FMT}\xspace,
  /molecules/102/.initial=\ce{FtBu{-}FMT}\xspace,
  /molecules/12/.initial=\ce{$4’$Cl{-}MPF}\xspace,
  /molecules/93/.initial=\ce{tBu{-}$4’$Cl{-}MPF}\xspace,
  /molecules/98/.initial=\ce{FtBu{-}$4’$Cl{-}MPF}\xspace
}

\newcommand{\molecule}[1]{\pgfkeysvalueof{/molecules/#1}}

\author{Ivan Tambovtsev}
\affiliation[University of Iceland]{Science Institute and Faculty of Physical Sciences, University of Iceland, 107 Reykjav\'{\i}k, Iceland}
\email{ivt3@hi.is}
\author{Hannes Jónsson}
\affiliation[University of Iceland]{Science Institute and Faculty of Physical Sciences, University of Iceland, 107 Reykjav\'{\i}k, Iceland}
\email{hj@hi.is}

\title{
Tuning Molecular Motors with Tert-Butyl and Fluorinated Tert-Butyl Groups
}

\begin{document}

\renewcommand*\tocentryname{TOC Graphic}
\begin{tocentry}
   \includegraphics[width = \textwidth]{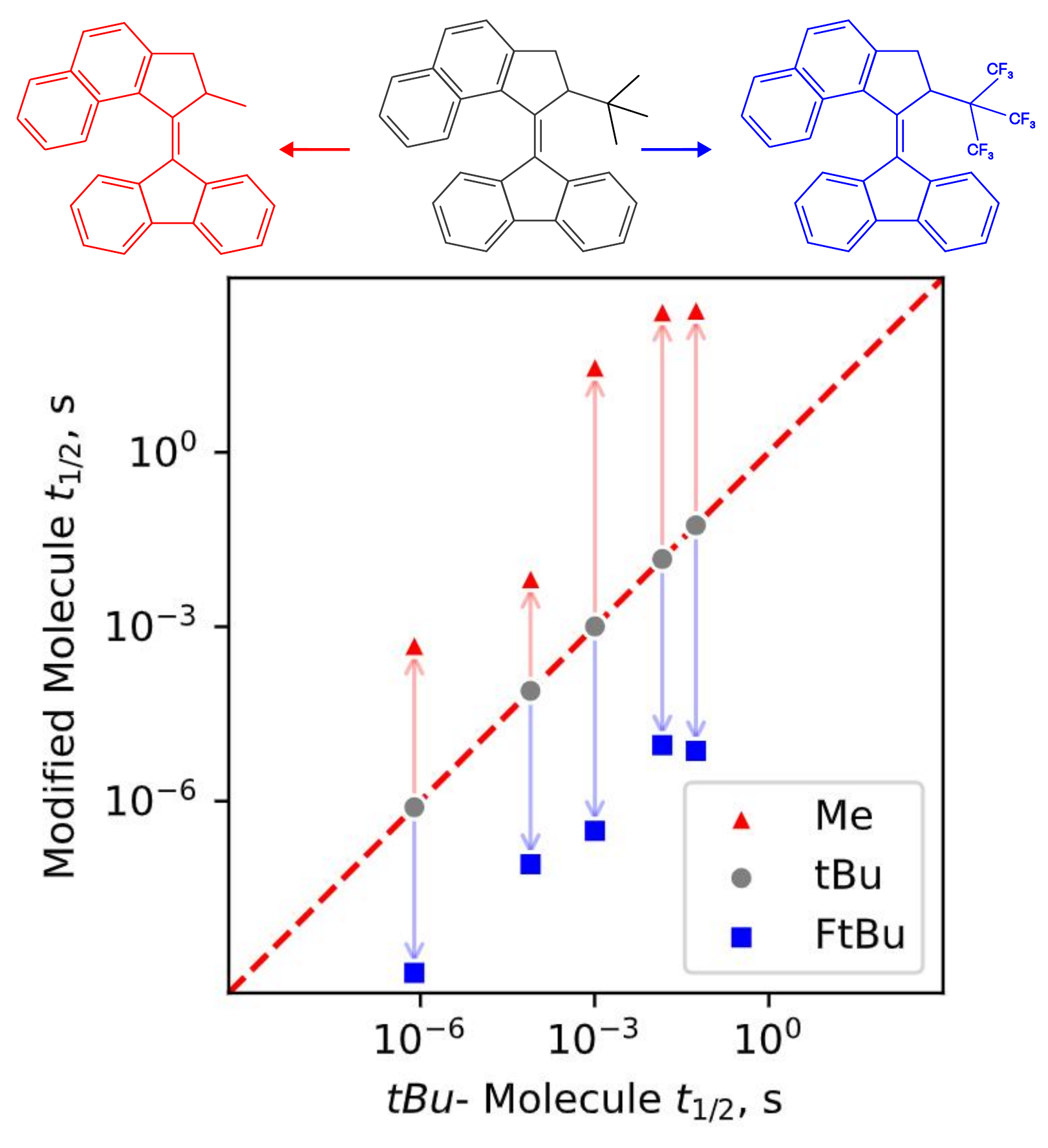}
\end{tocentry}

\begin{abstract} 
The effect of modifying substituents in the rotor group of five second generation molecular motors is estimated by theoretical calculations. The rotational speed is estimated by calculating the rate limiting step, the thermal helix inversion, as well as the competing backward transition using harmonic transition state theory with energy and atomic forces obtained from  density functional theory. First, a methyl group at the stereogenic center is replaced with a tert-butyl (tBu) group and the rotational speed is found to increase due to reduced lifetime of the metastable state. For two of the rotors, comparison can be made with experimental measurements and the calculated half-life is in close agreement. Secondly, the effect of substituting the nine hydrogen atoms in the tBu group with fluorine atoms is studied and this is found to increase the rotational rate further without significantly altering the molecular structure. The excitation wavelength of both the stable and metastable states is calculated and the separation of the absorption peaks is found to increase by the tBu substitution and even more so by the fluorinated tBu substitution, up to 40 nm. These findings can help develop a strategy for designing molecular motors with a rotational speed that best fits a given application.
\end{abstract}


\section{INTRODUCTION}

The development of light-powered molecular motors where a part of a molecule undergoes a full rotation relative to the rest has gained significant attention in recent years, with numerous designs demonstrating efficient and directed molecular motion~\cite{corraPhotoactivatedArtificialMolecular2023, baronciniPhotoRedoxDrivenArtificial2020, jeongMolecularSwitchesMotors2020}. These molecules typically feature a C=C double bond with one side designated as the ``rotor" and the other as the ``stator".
Upon photon absorption, the molecule undergoes an initial 90$^{\circ}$ rotation in the excited state. Upon relaxation back to the ground state, steric effects drive further rotation to a metastable state. A thermally activated transition, the thermal helix inversion (THI), then drives the molecule to a second stable state corresponding to a rotation of half a circle. Upon absorption of a second photon followed by THI, the full 360$^{\circ}$ rotation is completed.\cite{garcia-lopezLightActivatedOrganicMolecular2020, poolerDesigningLightdrivenRotary2021, rokeMolecularRotaryMotors2018, koumuraLightdrivenMonodirectionalMolecular1999}.

For efficient unidirectional operation, the rate of the forward THI must be greater than that of the reverse transition back to the stable state, which is referred to as thermal isomerization (TI). 
When the stator is symmetric, the stable states are structurally equivalent and the directionality is determined solely by the asymmetry of the energy landscape.

Tuning of the rotational rate to an optimal value 
is important
for various applications, for example systems where rotation at the nanoscale is translated into macroscopic actuation, such as in liquid crystalline networks or photonic materials~\cite{houPhototriggeredComplexMotion2022, lanAmplifyingMolecularScale2022, houPhotoresponsiveHelicalMotion2021, sunLightDrivenSelfOscillatingBehavior2021}. Furthermore, the efficiency is reduced if the metastable state absorbs a photon so a sufficient spacing between the absorption peaks of the stable and metastable states is needed.
Extensive efforts have focused on structural modifications that optimize light absorption and enhance the rate of the THI while still ensuring the TI is slow enough.\cite{poolerDesigningLightdrivenRotary2021, vicarioControllingSpeedRotation2005} 
Among such modifications is the replacement of hydrogen atoms by fluorine to adjust electronic and steric properties.
\cite{huangLongLivedSupramolecularHelices2018, blegerOFluoroazobenzenesReadilySynthesized2012}.

It has been demonstrated experimentally that fluorination can have contrasting effects: a CF\textsubscript{3} substitution can make rotation faster, while substitution of H by F at the stereogenic center can have the opposite effect, raising the energy barrier for the THI and thereby slowing down the rotation.\cite{stackoFluorineSubstitutedMolecularMotors2017}
In a previous computational study, we demonstrated that 
calculations of minimum energy paths (MEPs) and application of harmonic transition state theory (HTST)
with energetics obtained from density functional theory (DFT) give estimates of the thermal rate constants 
that are in close agreement with experimental data for a wide range of motors.\cite{tambovtsevFineTuningRotational2025} 
For several rotors we found that
substitution of a CH\textsubscript{3} group by a CF\textsubscript{3} group at the stereogenic center accelerates the THI by raising the energy of the metastable state without significantly affecting the transition structure (TS).
Targeted fluorination strategies can significantly enhance performance of molecular motors and thereby extend their utility in the various possible applications such as soft robotics, responsive materials, and chiral optics~\cite{yangLighttriggeredModulationSupramolecular2023, kimPhotoresponsiveChiralDopants2019, orlovaRevolvingSupramolecularChiral2018}.

\begin{figure}[H]
\includegraphics[width = \textwidth]{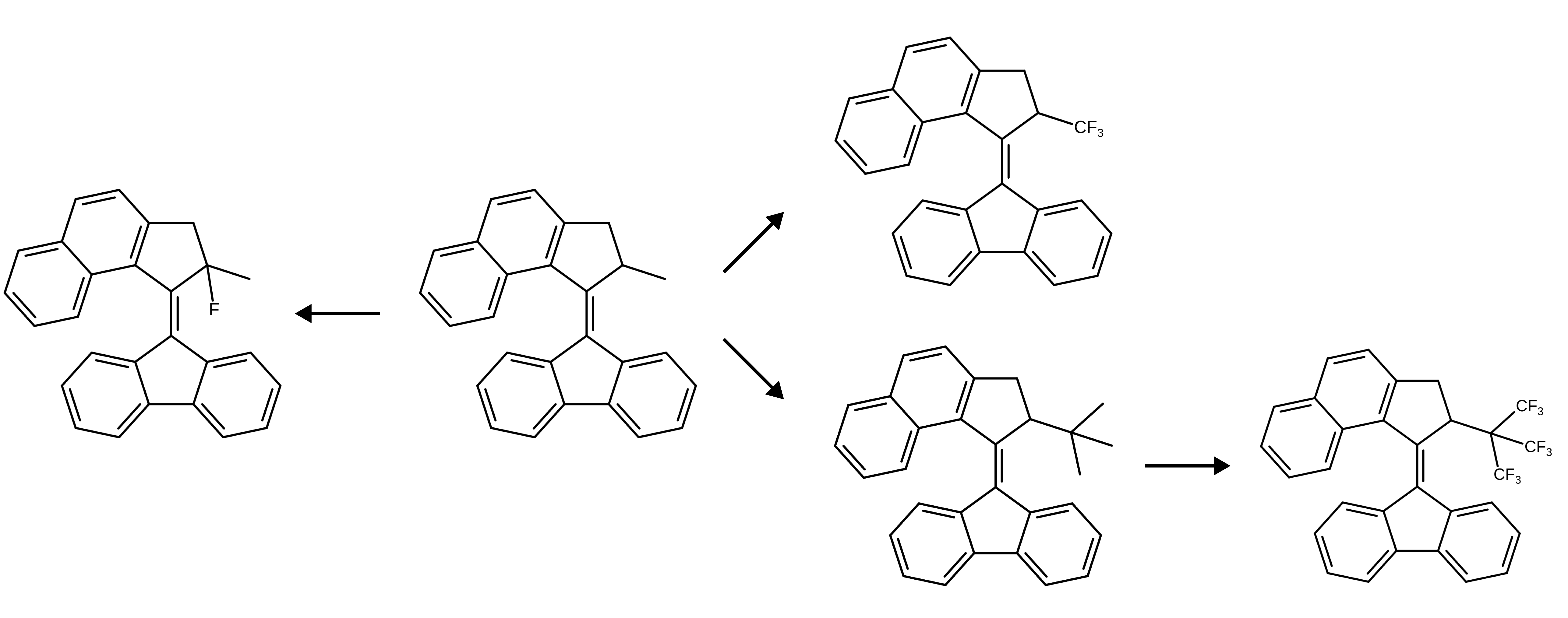}
\caption{Examples of structural modifications of a second generation molecular motor. Starting from the methyl-substituted rotor (center), the molecule is modified either by fluorination of the stereogenic center (left), 
substitution of the methyl group with a trifluoromethyl group (top), 
replacement of the methyl group with a tert-butyl group (bottom-left), and full fluorination of the tert-butyl group (bottom-right).
}
\label{fig:transition}
\end{figure}

In the present study, we calculate 
the effect of substituting a methyl group by a tert-butyl group or a fully fluorinated tert-butyl group.
Such structural modifications, examples of which are shown in figure~\ref{fig:transition}, can give control over the rotational speed.
The absorption wavelengths of the stable and metastable states are also estimated and the substitutions are found to increase the separation between the absorption peaks, thereby improving efficiency.

%


%



\section{METHODS}

The minimum energy path for the THI and TI transitions from the metastable state 
is calculated using the climbing image nudged elastic band (CI-NEB) method \cite{millsReversibleWorkTransition1995,henkelmanClimbingImageNudged2000,henkelmanImprovedTangentEstimate2000},
as implemented in the ORCA software\cite{neeseORCAProgramSystem2012,neeseSoftwareUpdateORCA2022} using energy weighted springs\cite{asgeirssonNudgedElasticBand2021}.
The initial path is generated using the sequential image dependent pair potential (S-IDPP) method\cite{schmerwitzSaddlePointSearch2024}.
The locations of the images along the path are converged to a tolerance of $2.5\cdot 10^{-3}$\,au and $5\cdot 10^{-3}$\,au\ in the RMS and MAX of the force perpendicular to the local tangent to the path, respectively, with one order of magnitude tighter tolerance on the climbing image. CI-NEB is followed by a first-order saddle point search (SPS) with 
the CI as the initial guess,\cite{asgeirssonNudgedElasticBand2021} and is converged to a tolerance of $5\cdot 10^{-6}$\,au, $3\cdot 10^{-4}$\,au, $10^{-4}$\,au, $4\cdot 10^{-3}$\,au, and $2\cdot 10^{-3}$\,au\ for the switch from CI-NEB to SPS, the MAX and RMS of the gradient, and the MAX and RMS of the optimization step, respectively. 

The DFT calculations make use of the B3LYP hybrid functional approximation\cite{leeDevelopmentColleSalvettiCorrelationenergy1988, beckeDensityfunctionalExchangeenergyApproximation1988, beckeDensityFunctionalThermochemistry1993} 
and a linear combination of atomic orbitals with the 6-31G(d,p) basis set~\cite{hehreSelfConsistentMolecular2003, weigendAccurateCoulombfittingBasis2006}. 
The B3LYP functional is chosen for its proven performance in a wide range of molecular systems, and the 6-31G(d,p) basis set is selected for its balance between computational efficiency and accuracy. 
The structure of the stable and metastable states of all molecular motors is optimized to a tolerance of $10^{-4}$\,au, $3\cdot 10^{-4}$\,au, $2\cdot 10^{-3}$\,au, and $4\cdot 10^{-3}$\,au\ for the RMS and MAX of the gradient and the RMS and MAX of the optimization step, respectively. 
Atom coordinates of the local minima and TS as well as scripts specifying tolerances are provided in the data stored and openly provided on Zenodo.

The rate constants for the THI and TI are calculated using the harmonic approximation to transition state theory\cite{wignerTransitionStateMethod1938,vineyardFrequencyFactorsIsotope1957}
\begin{equation}
k_{\rm HTST} = \frac{\prod_i^{3N} \nu_i^{\rm min}}{\prod_i^{3N-1} \nu_i^\ddagger} 
    \exp \left[ -\left(E^\ddagger - E^{\rm min}\right)/k_{\rm B} T \right],
    \label{eqn:htst-rate}
\end{equation}
where $\nu_i^{\rm min}$ and $\nu_i^\ddagger$ refer to vibrational frequency, and $E^{\rm min}$ and $E^\ddagger$ refer to the energy of the initial state minimum and first order saddle point, respectively. 
The vibrational analysis is furthermore used to confirm that the stable and metastable structures are local minima on the energy surface and the TSs correspond to first-order saddle points. 
The half-life of the metastable state is obtained from the THI rate constant as $\tau = \ln{2}/k_{\rm HTST} $. 
Since the rate of the TI turns out to be so much smaller than the THI, its effect can be neglected in the half-life calculation.

The absorption wavelengths are obtained with linear-response TDDFT within the adiabatic approximation. All calculations are performed with the ORCA 5 software\cite{neeseORCAProgramSystem2012, neeseSoftwareUpdateORCA2022}. 
The data was extracted using ChemParse.


\section{RESULTS AND DISCUSSION}

\begin{figure}[H]
\includegraphics[width=0.8\textwidth]{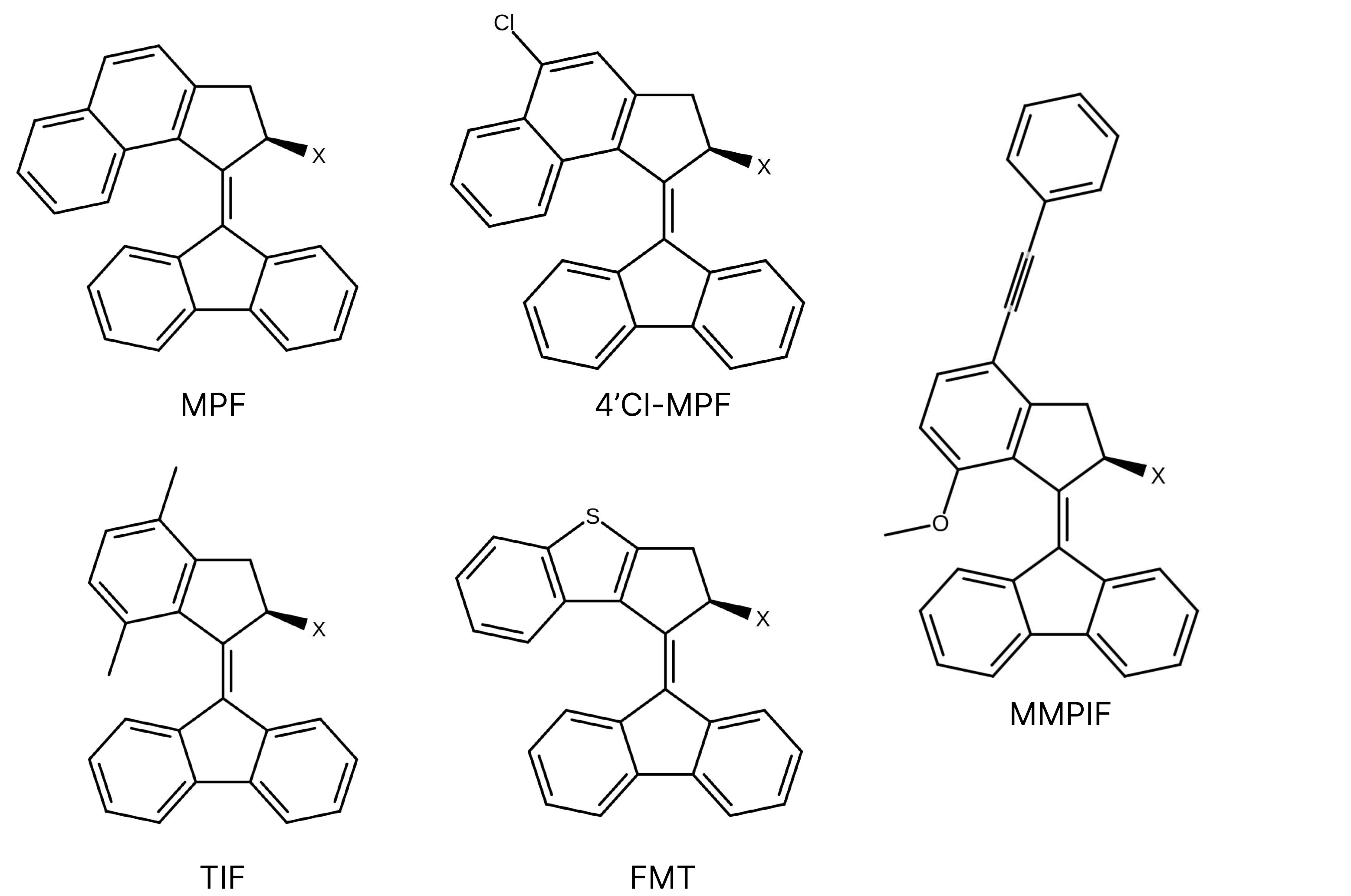}
\caption{
The molecular motors investigated in the present study to assess the effect of modifications at site X. In the base molecules, X is a CH\textsubscript{3} group (methyl), while in the modified molecules, X is replaced by a tert-butyl or fluorinated tert-butyl group.
}
\label{fig:molecules}
\end{figure}

To assess the impact of structural modifications on the rotational rate of second-generation molecular motors, we selected a set of five well-characterized base molecules, illustrated in figure~\ref{fig:molecules}. They are representative of the common second-generation motor designs, featuring 
a five-membered ring in the rotor connected to a fluorene stator by a C=C double bond. They have all been synthesized and experimentally studied~\cite{vicarioFineTuningRotary2006, pollardRedesignLightdrivenRotary2008, pollardEffectDonorAcceptor2008, cnossenTrimerUltrafastNanomotors2009, garcia-lopezLightActivatedOrganicMolecular2020}, and were chosen for this study because they exhibit a diverse range of structural features and rotational properties, providing a basis for evaluating systematic trends in the effect of modifications. 
For each of these five base molecules, illustrated in figure~\ref{fig:molecules}, three variants at the substitution site X were studied: the original methyl group (CH$_3$), a tert-butyl group (tBu), and a fully fluorinated tert-butyl group (FtBu). 

\begin{figure}[H]
  \includegraphics[width = 0.5\textwidth]{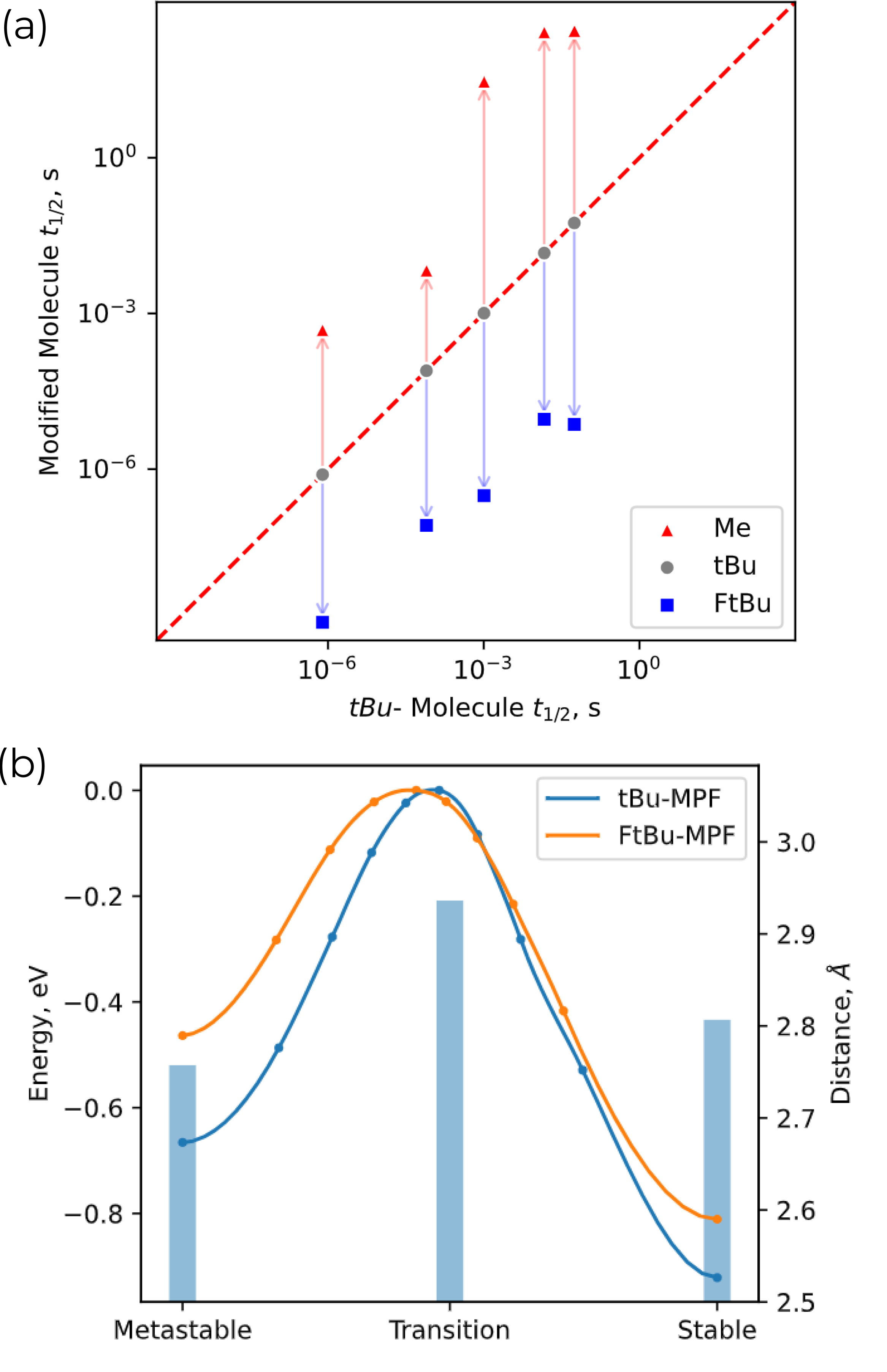}
  \caption{
  (a) Calculated half-life of the metastable state of the molecular motors and the effect of substituting the tert-butyl (tBu) group at site X with either a methyl group, Me (red) or a fully fluorinated tert-butyl group, FtBu (blue). 
  The FtBu substitution consistently increases the forward rotation rate, while the Me substitution decreases it.  
  (b) Minimum energy paths for the THI of \molecule{3} and \molecule{10}. The zero of energy is set to that of the transition structure, TS. The bars and corresponding axis to the right show the average distance between the hydrogen atoms in the tBu group on the rotor of \molecule{3} to atoms in the stator that are within 2.9~\AA{} in the stable, TS, and metastable state. 
  This serves as a measure of the steric hindrance, which is smaller in the TS than in the stable and metastable states.
  The substitution of tBu by FtBu thereby reduces the energy barrier for the THI.
  }
  \label{fig:rates-alt}
\end{figure}

Figure~\ref{fig:rates-alt}(a) illustrates the calculated half-life of the metastable state of the rotors with methyl, tBu and FtBu at site X. The changes in the half-life are presented in a way that the rotors start with tBu and this group is then replaced by methyl or FtBu.
The effect of the substitutions is in the same direction for all five rotors. The methyl group gives the longest half-life and thereby slowest rotation, while the FtBu gives the shortest half-life and fastest rotation.
In all cases, the back transition, TI, is slower than the THI. 
The calculated values of the half-life as well as the rate constants are given in table 1. 
For two of the tBu molecular motors, experimentally determined values of the half-life have been reported
~\cite{vicarioFineTuningRotary2006, pollardRateAccelerationLightDriven2007} and our calculated values are in close agreement, within a factor of three:  
5.7$\times$10$^{-3}$ s vs. 1.46$\times$10$^{-2}$ s for \molecule{3}, and 2.6$\times$10$^{-3}$ s vs. 1.0$\times$10$^{-3}$ s for \molecule{9}.

Fluorination of the tBu group to form FtBu 
does not
significantly alter the molecular structure compared to the tBu analogues. The acceleration of the rotation 
arises from increased steric hindrance by the fluorine atoms in the stable and metastable states.
Figure ~\ref{fig:rates-alt}(b) shows the calculated MEP for rotors \molecule{3} and \molecule{10}. 
Clearly, the activation energy for the THI is reduced when tBu is replaced by FtBu and the rotational speed thereby increased.

\begin{table}[H]
\centering
\begin{tabular}{|c|c|c|c|c|c|c|}
\hline
\textbf{Molecule} & \textbf{$\lambda_{\text{stable}}$, nm} & \textbf{$\lambda_{\text{meta}}$, nm} & \textbf{$k^{THI},\ \mathrm{s}^{-1}$} & \textbf{$k^{TI},\ \mathrm{s}^{-1}$} & \textbf{$t_{1/2},\ \mathrm{s}$} & \textbf{$t_{1/2}^{Exp},\ \mathrm{s}$} \\
\hline
\molecule{0} & 379 & 410 & $2.60 \times 10^{-3}$ & $8.34 \times 10^{-4}$ & $2.02 \times 10^{2}$ & $1.90 \times 10^{2}$ \ \ \cite{vicarioFineTuningRotary2006} \\
\hline
\molecule{3} & 371 & 437 & $4.75 \times 10^{1}$ & $7.99 \times 10^{-3}$ & $1.46 \times 10^{-2}$ & $5.73 \times 10^{-3}$ \ \ \cite{vicarioFineTuningRotary2006} \\
\hline
\molecule{10} & 368 & 442 & $7.59 \times 10^{4}$ & $6.26 \times 10^{4}$ & $5.00 \times 10^{-6}$ & - \\
\hline
\molecule{7} & 340 & 367 & $2.32 \times 10^{-2}$ & $7.93 \times 10^{-7}$ & $2.99 \times 10^{1}$ & $1.50 \times 10^{1}$ \ \ \cite{pollardRedesignLightdrivenRotary2008} \\
\hline
\molecule{9} & 336 & 395 & $6.85 \times 10^{2}$ & $2.61 \times 10^{-3}$ & $1.01 \times 10^{-3}$ & $2.90 \times 10^{-3}$ \ \ \cite{bauerTuningRotationRate2014} \\
\hline
\molecule{11} & 332 & 405 & $2.24 \times 10^{6}$ & $1.09 \times 10^{-1}$ & $3.09 \times 10^{-7}$ & - \\
\hline
\molecule{12} & 388 & 419 & $2.44 \times 10^{-3}$ & $1.84 \times 10^{-3}$ & $1.62 \times 10^{2}$ & $2.88 \times 10^{2}$ \ \ \cite{pollardEffectDonorAcceptor2008} \\
\hline
\molecule{93} & 380 & 446 & $1.25 \times 10^{1}$ & $1.68 \times 10^{-2}$ & $5.54 \times 10^{-2}$ & - \\
\hline
\molecule{98} & 372 & 454 & $9.52 \times 10^{4}$ & $4.39 \times 10^{-1}$ & $7.28 \times 10^{-6}$ & - \\
\hline
\molecule{50} & 349 & 389 & $1.42 \times 10^{3}$ & $8.78 \times 10^{-6}$ & $4.88 \times 10^{-4}$ & $1.60 \times 10^{-4}$ \ \ \cite{cnossenTrimerUltrafastNanomotors2009} \\
\hline
\molecule{95} & 345 & 394 & $8.83 \times 10^{5}$ & $5.54 \times 10^{-5}$ & $7.85 \times 10^{-7}$ & - \\
\hline
\molecule{100} & 334 & 399 & $6.22 \times 10^{8}$ & $5.21 \times 10^{-4}$ & $1.11 \times 10^{-9}$ & - \\
\hline
\molecule{68} & 347 & 365 & $1.02 \times 10^{2}$ & $1.12 \times 10^{-6}$ & $6.80 \times 10^{-3}$ & $7.00 \times 10^{-2}$ \ \ \cite{garcia-lopezLightActivatedOrganicMolecular2020} \\
\hline
\molecule{97} & 345 & 384 & $8.80 \times 10^{3}$ & $3.04 \times 10^{-6}$ & $7.88 \times 10^{-5}$ & - \\
\hline
\molecule{102} & 347 & 397 & $8.39 \times 10^{6}$ & $2.18 \times 10^{-3}$ & $8.26 \times 10^{-8}$ & - \\
\hline
\end{tabular}
\caption{Absorption wavelengths for the stable and metastable isomers, corresponding to the excitation that triggers photoinduced rotation, calculated rate constants for the forward (THI) and backward (TI) thermal reactions, and the corresponding theoretical and experimental half-lives.}
\label{table:spectroscopy}
\end{table}

To better understand the origin of this acceleration, the molecular structure along the MEPs is analyzed by focusing on the steric interactions between the rotor and stator. Specifically, we use the same metric as in our previous studies of fluorination
\cite{tambovtsevFineTuningRotational2025}:
 the average distance between the hydrogen atoms of the rotor in the tBu group and the atoms of the stator that lie within 2.9~\AA, which corresponds to the sum of the van der Waals radii for carbon and hydrogen atoms.
 This serves as an indicator of steric hindrance. 
 Figure~\ref{fig:rates-alt}(b) illustrates the value of this average distance for the stable, TS and metastable states.
 The TS turns out to have the largest value and thereby smaller steric hindrance, while the metastable state has the smallest value and largest steric hindrance. 
 The replacement of tBu with FtBu therefore raises the energy of the metastable state more than the TS, and thereby lowers the 
 energy barrier for the THI.

\begin{table}[H]
\centering
\begin{tabular}{|c|c|c|c|}
\hline
Molecule &
$\bar{d}_{\text{stable}}$ &
$\bar{d}_{\text{TS}}$ & $\bar{d}_{\text{meta}}$  \\
\hline
\molecule{3} & 2.81 & 2.94 & 2.76 \\
\molecule{9} & 2.82 & 2.89 & 2.74 \\
\molecule{93} & 2.82 & 2.91 & 2.75 \\
\molecule{95} & 2.78 & 2.83 & 2.78 \\
\molecule{97} & 2.61 & 2.84 & 2.73 \\
\hline
\end{tabular}
\caption{Average rotor–stator proximity (in \AA) for tBu-substituted motors in stable state, transition structure and metastable state.}
\label{table:Distance}
\end{table}

Table~\ref{table:Distance} shows the value of this measure of steric hindrance for all five tBu-substituted molecules. The same trend is seen.
This measure of the average rotor–stator distance is relatively larger in the TS and shortest in the metastable state, indicating that steric hindrance is more pronounced there. 
As a result, after the replacement of the hydrogen atoms in the tBu group with fluorine, the energy of the metastable state is increased relative to the TS.

\begin{figure}[H]
  \includegraphics[width = 0.5\textwidth]{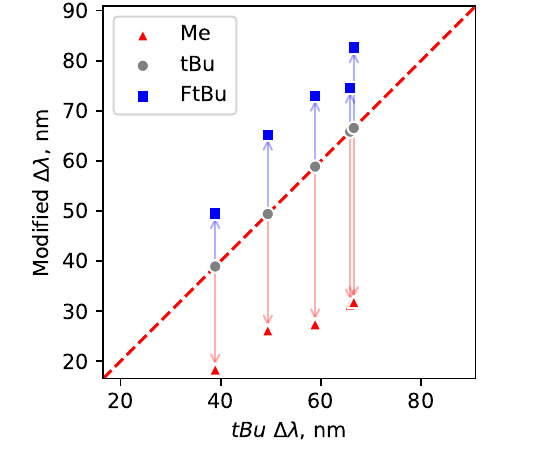}
  \caption{Calculated difference in the excitation wavelength of the stable and metastable states, $\Delta \lambda$, for the five motors with the various substituents. The methyl-substituted motors, Me, consistently give the smallest gaps (red), 
  tert-butyl, tBu, increases the separation (gray), and fully fluorinated tert-butyl, FtBu, increases the gap even further (blue).}
  \label{fig:gap}
\end{figure}

Figure~\ref{fig:gap} illustrates the difference in the absorption wavelength of the metastable and stable states, $\Delta \lambda$, when the tBu
is replaced by methyl or FtBu.
The results reveal a systematic trend: the rotors with the methyl group have the smallest gap, those with tBu have a larger gap by 20 to 30 nm, 
and the largest gap is obtained for the rotors with FtBu, increased by as much as 42~nm compared with rotors where a methyl group is at the X site.
This increasing spectral separation enhances selective photoactivation by 
minimizing the risk of unwanted photoreversal that can occur if the metastable state is excited. 
The complete set of calculated absorption wavelengths is provided in Table~\ref{table:spectroscopy}).


\section{CONCLUSIONS}
The substitution of a methyl group with a tBu group or a FtBu group at the stereogenic center of five second generation molecular motors is found to increase the rotational speed as well as the separation of the absorption peaks for the stable and metastable states. The trend is the same for all the rotors and the FtBu substitution gives a larger effect than the tBu substitution. 
The reason for the increased rotational speed is larger steric hindrance in the metastable state than in the transition structure and thereby lower energy barrier for the THI.
Where experimental data is available, the agreement with the calculations presented here based on calculations of the MEP and rates using HTST with energetics coming from DFT/B3LYP is good. The calculations are expected to have a predictive value and can be used to guide the fine tuning of rotor molecules.

\begin{acknowledgement}

This work was funded by the Icelandic Research Fund (grants 239970 and 2511544).
We thank Gianluca Levi for helpful discussions.
The calculations were carried out at the IREI HPC facility at the University of Iceland.

\end{acknowledgement}

\section{Data Availability Statement}
The data supporting the findings of this work are available for download at Zenodo.

\bibliography{zotero_tambovtsev}

\end{document}